\documentstyle[12pt,a4]{article}
\def\3{\ss}
\renewcommand{\theequation}{\thesection.\arabic{equation}}

\newcommand{\gap}{\stackrel{>}{\sim}}
\newcommand{\chibarchi}{\langle\bar{\chi}\chi\rangle}  % chibarchi
\newcommand{\figurebox}[2]{\fbox{\vbox
 to#2cm{\hbox to #1cm{\hfil} \vfil}}}
\begin{document}
%
%-----------------------------------------------------------------------
% The lines below are necessary in order to enumerate the equations
% according to the sections where they are.
\makeatletter
\@addtoreset{equation}{section}
\makeatother
\renewcommand{\theequation}{\thesection.\arabic{equation}}
%-----------------------------------------------------------------------
\title{Is the Chiral Phase Transition in Non-Compact  \\
Lattice QED Driven by Monopole Condensation?}
\author{M. G\"ockeler$^{1,2}$, R. Horsley$^{2}$,
P.E.L. Rakow$^3$ and
              G. Schierholz$^{2,4}$
 \\[2em]
$^1$ Institut f\"ur Theoretische Physik, RWTH Aachen, \\
Sommerfeldstra\3e, W-5100 Aachen, Germany\\[0.5em]
$^2$Gruppe Theorie der Elementarteilchen,\\
H\"ochstleistungsrechenzentrum HLRZ, \\
c/o KFA, Postfach 1913, W-5170 J\"ulich, Germany\\[0.5em]
$^3$ Institut f\"ur
Theoretische Physik, Freie Universit\"at Berlin, \\
Arnimallee 14,
W-1000 Berlin 33, Germany\\[0.5em]
$^4$ Deutsches Elektronen-Synchrotron DESY, \\
Notkestra\3e 85, W-2000 Hamburg 52, Germany}
  \date{ \vspace{-16.5cm} {\rm
    \begin{flushleft} DESY 93-025 \\
                      FUB-HEP 6/93 \\
                      HLRZ 93-12 \\[0.5em]
                      March 1993 \end{flushleft}} \vspace{12.5cm}}
\maketitle
\begin{abstract}

We investigate the recent conjecture that the chiral phase transition
in non-compact lattice QED is driven by monopole condensation. The
comparison of analytic and numerical results shows that we have a
quantitative understanding of monopoles in both the quenched and
dynamical cases. We can rule out monopole condensation.

\end{abstract}

\newpage

\section{Introduction}

In a series of papers
%\cite{GHLRSSW1,Sch1,GHLRSSW2,Rakow1,Rakow2,Sch2,%
%Hors1,Go1,Sch3,GHRSS1}
 [1-10]
we have investigated strongly coupled QED, both on the lattice
in the non-compact formulation and in the continuum
using Schwinger-Dyson equations. The strong coupling region is of
interest because of the existence of a second order chiral phase
transition. This implies a continuum limit, and the interesting
question is whether or not it describes an interacting theory. Our
calculations of the renormalized charge, $e_R$, and fermion mass,
$m_R$, demonstrated that whenever $m_R$ goes to zero in lattice units
(i.e. the ultraviolet cut-off is removed) then $e_R$ goes to zero.
This suggests that the theory is non-interacting in the continuum
limit in accordance with the general belief that non-asymptotically
free theories are trivial. It is encouraging that the
two approaches, namely lattice and Schwinger-Dyson,
agree with each other. Further support of this picture
comes from other authors~\cite{other}.

However, this picture has been queried by Hands, Koci\'c, Kogut and
collaborators
%\cite{Ko1,Ko2,KoN,Ko3,Ko4,Ko5,Ko6,Ko7},
[12-20]
who investigated the behaviour of magnetic monopoles
near the phase transition.
Using a new monopole `order parameter' first introduced
by Hands and Wensley~\cite{H&W}, they conclude that monopoles condense
in the chirally broken phase. The occurrence of this proposed second
order monopole phase transition is important because such a transition
would imply the existence of monopoles in the lattice model's
continuum limit. This would cast doubt on conclusions about continuum
physics drawn from lattice calculations, as monopoles are presumably
absent in QED. Furthermore, dual superconductivity
and charge confinement are to be expected
whenever monopoles condense~\cite{Ko3}.
This is in conflict with our picture~\cite{GHRSS1} where we have found
free electrons and massless photons in the broken phase.

In the strong coupling limit $\beta \rightarrow 0$, or the limit when
we have a large number of flavours, the action is dominated by the
fermion determinant, which is a compact object in the sense that it
only depends on the compactified link variables
 ${\rm e}^{i A_{\mu}}$, $A_{\mu}$ being the gauge
field. It was found that the compact U(1)
Wilson action has a first order phase transition~\cite{DeGrand}
 which is driven by
monopole condensation~\cite{DeGrand,first}. Hence, it is
conceivable that monopoles play a role in the non-compact
case as well. Indeed, simulations with very
large numbers of fermions~\cite{largeN} suggest that the phase
transition becomes first order.

In this paper we shall investigate the relevance of monopoles for the
phase transition. We solve the quenched case analytically~\cite{Rakow4}
and look at the dynamical fermion case numerically. We find no
evidence of monopole condensation.
\vspace{0.5cm}

%\newpage

\section{Lattice monopoles}

The action for non-compact lattice QED with
 dynamical staggered fermions is
\begin{equation}
S = \frac{\beta}{2} \sum_{x,\mu < \nu} F^2_{\mu \nu}(x)
+\sum_{x, y} \bar{\chi}(x) [D_{x , y} + m \delta_{x , y}]\chi(y),
\end{equation}
with
\begin{eqnarray}
F_{\mu \nu}(x) &=& \Delta_{\mu} A_{\nu}(x) - \Delta_{\nu} A_{\mu}(x), \\
\displaystyle
D_{x , y} &=& \frac{1}{2} \sum_{\mu} (-1)^{x_1 + \cdots
+ x_{\mu - 1}} [{\rm e}^{i A_{\mu}(x)} \delta_{y , x+\hat{\mu}}
- {\rm e}^{-i A_{\mu}(y)} \delta_{y , x-\hat{\mu}}],
\end{eqnarray}
where $\Delta_{\mu}$ is the lattice forward derivative and
$\beta = 1/e^2$.

To define monopoles~\cite{DeGrand,H&W} we decompose
$F_{\mu \nu}$ into an integer valued string field $N_{\mu \nu}$ and a
compact field $f_{\mu \nu}$ which lies in the range $(-\pi,\pi]$:
\begin{equation}
F_{\mu \nu} = 2 \pi N_{\mu \nu} + f_{\mu \nu}.
\end{equation}
The Bianchi identity tells us that $F_{\mu \nu}$ summed over any closed
surface always gives zero. This does not apply to the $N_{\mu \nu}$
and $f_{\mu \nu}$ fields separately. This allows the common
definition of a monopole current
\begin{equation}
M_{\mu}(x) = \frac{1}{4 \pi} \epsilon_{\mu \nu \rho \sigma}
\Delta_{\nu} f_{\rho \sigma}(x+\hat{\mu}) = -\frac{1}{2}
\epsilon_{\mu \nu \rho \sigma} \Delta_{\nu}
N_{\rho \sigma}(x+\hat{\mu}),
\label{Mu}
\end{equation}
which lives on the elementary cubes of the lattice (or
 equivalently on the links of the dual lattice).
 Equation~(\ref{Mu}) shows that
 a string can only end on a monopole or antimonopole.
 Each component of $M_\mu$ can take the values $0, \pm 1, \pm 2$.
 The current $M_{\mu}$ is conserved: $\bar\Delta_{\mu} M_{\mu}(x) = 0$,
 where $\bar\Delta_\mu$ is the lattice backward derivative.

Later on we shall be interested in the monopole susceptibility. This
is defined by~\cite{Cardy}
\begin{equation}
\chi_m = \frac{1}{6} \sum_x \langle f_{\mu \nu}(x)
f_{\mu \nu}(0) \rangle
= \frac{4 \pi^2}{6} \sum_x \langle N_{\mu \nu}(x) N_{\mu \nu}(0)
\rangle ,
\label{chif}
\end{equation}
where we have made use of the fact that $\sum_x F_{\mu \nu}(x) = 0$.
 In the infinite volume limit further manipulations lead to
 the equivalent form
\begin{equation}
\chi_m = -\frac{4 \pi^2}{12} \sum_{x} \langle x^2 M_{\mu}(x)
M_{\mu}(0) \rangle \, .
\label{chiM}
\end{equation}
The physical interpretation of
eq.~(\ref{chiM}) is that
it measures the fluctuations of the total dipole
moment, whereas eq.~(\ref{chif})
can be regarded as the residue of the photon
pole in the compact photon propagator.  We have used eq.~(\ref{chif})
for our measurements to avoid ambiguities in defining $x^2$ on
a finite lattice. If the monopoles condense we
would expect $\chi_m$ to diverge.
\begin{figure}[b]
\vspace{13.5cm}
 \noindent \special{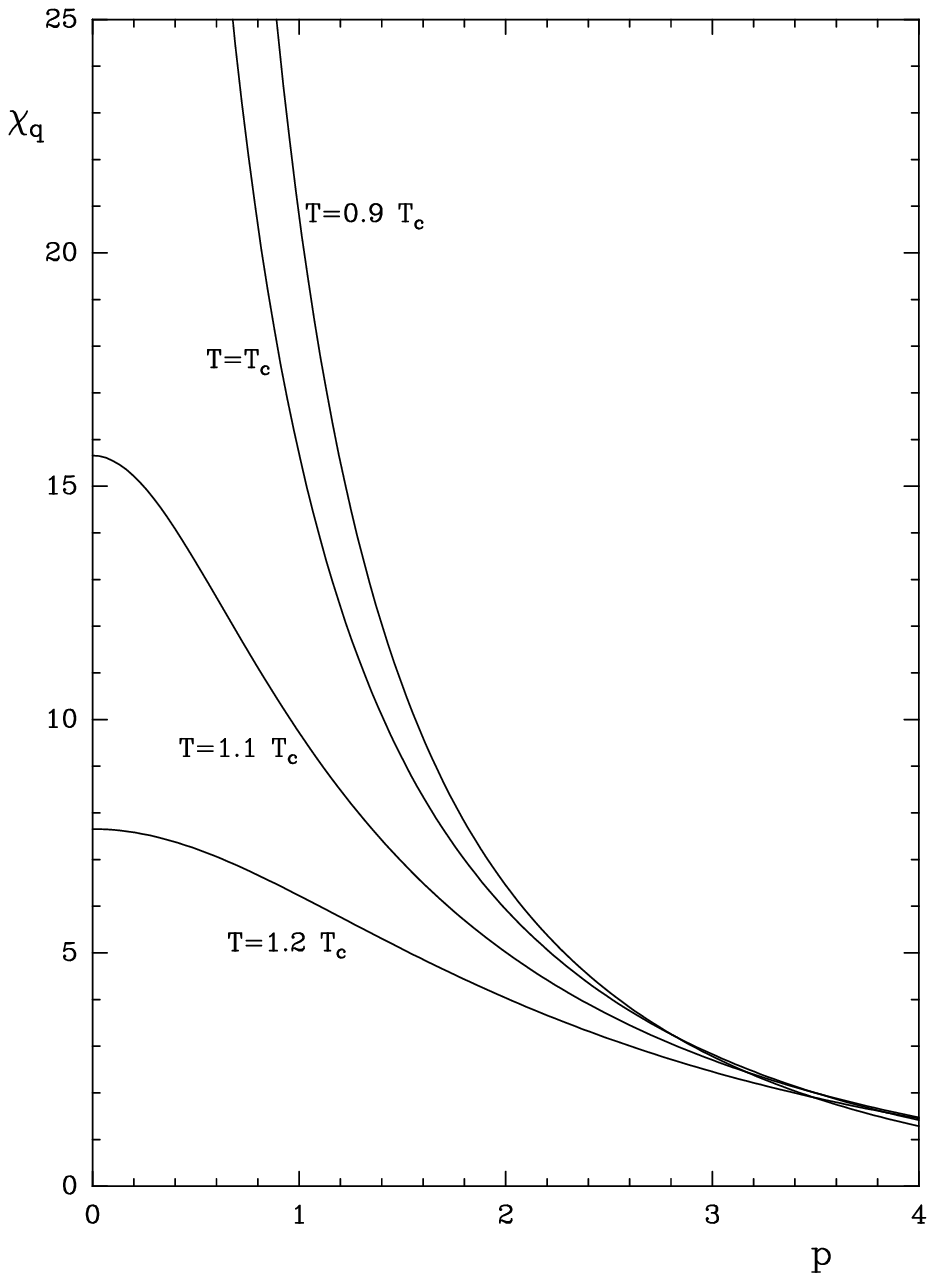}
 \hspace{2cm}\parbox{12cm}{\caption{  The susceptibility (2.8) for an
  ideal Bose-Einstein gas, at various temperatures above
 and below the condensation temperature $T_c$.} }
\end{figure}

At a second order phase transition fluctuations of physical quantities
become large and should show up in other quantities as well. We have
looked at fluctuations in monopole charge. The corresponding
susceptibility is
\begin{equation}
\chi_q(p) = \frac{1}{4} \sum_{x, \mu} \langle M_{\mu}(x)
M_{\mu}(0) \rangle {\rm e}^{i p \cdot x} \, .
\label{chiq}
\end{equation}
 It has long been known that when Bose condensation occurs
 the tendency of bosons to occupy the same state causes
 long range density correlations, which in turn leads to
 a divergence of $\chi_q$ at small momentum. For an example
 see `Distance Correlations and Bose-Einstein Condensation'
 \cite{London} which finds that in a Bose gas with a condensation
 temperature $T_c$ the density-density correlation
 drops off exponentially if $T>T_c$ (giving
 a finite $\chi$ at $p=0$), drops off like $r^{-2}$ at $T_c$
 (giving a $\chi$ diverging like $p^{-1}$) and drops off like
 $r^{-1}$ below $T_c$ (giving a $\chi$ diverging like $p^{-2}$).
 These behaviours are illustrated for the case of the ideal
 Bose gas in fig.~1. We see that condensation gives a spectacular
 signal.

  One subtlety is that for periodic boundary conditions
 $\chi_q(p)$ vanishes for $p = 0$
 on finite lattices and so one must extrapolate to zero momentum.
 (Note that the limits have to be taken in the correct order:
 volume goes to infinity before momentum goes to zero.)

Other quantities we shall also look at are the monopole density
\begin{equation}
\rho = \frac{1}{4 V} \sum_{x, \mu} |M_{\mu}(x)|
\end{equation}
($V$: lattice volume) and the string density \footnote{The normalization
is chosen to agree with ref.~\cite{H&W}.}
\begin{equation}
\sigma = \frac{1}{4 V} \sum_{x, \mu < \nu} |N_{\mu \nu}( x)| .
\label{sigma}
\end{equation}
We would expect that these quantities
 would show non-analytic behaviour
(though not necessarily a divergence)
 around a phase transition.
\vspace{0.5cm}

%\newpage

\section{Analytic results}

The authors of~\cite{H&W,Ko5} claim that even in quenched
non-compact QED there is a phase transition at which
the magnetic monopoles condense.
Because in the quenched case the action is Gaussian, we can derive
analytic formulae for most quantities~\cite{Rakow4}.
 The simplest
quantity is the string density. The probability distribution for a
single $F_{\mu\nu}$ field is a Gaussian:
\begin{equation}
\Psi(F) = \pi^{-\frac{1}{2} } \beta^{\frac{1}{2}}\,
         {\rm e}^{-\beta F^2} .
\label{psi1}
\end{equation}
The distribution is completely determined because we know the width
$\langle F^2 \rangle = 1/(2\beta)$.
 This gives (cf. eq.~(\ref{sigma}))
\begin{equation}
\begin{array}{ll}
\displaystyle
\sigma(\beta) &
\displaystyle   = \frac{3}{2} \langle |N(F)| \rangle \\ [1.2em]
\displaystyle
              &   \displaystyle
                = \frac{3}{2} \int_{-\infty}^{\infty} d F \,
                  \Psi(F)\, |N(F)| \\ [1.2em]
\displaystyle
              & \displaystyle = \frac{3}{2} \sum_{n=0}^{\infty}
       {\rm erfc}\left((2n+1)\pi \beta^{\frac{1}{2}}\right) \, .
\end{array}
\label{sigmaseries}
\end{equation}
 On a finite lattice with volume $V$ the only change is that the
 width of the Gaussian is reduced to
 $\langle F^2 \rangle = (V-1)/(2V \beta)$. This means that $\sigma$
 on a finite lattice can be found by making the replacement
 $\beta \to \beta V/(V-1)$ in eq.~(\ref{sigmaseries}).
 One sees that finite size effects
are already negligible on rather small lattices (which holds for all
the other quantities in this section as well).

The next quantity of interest is the monopole density. To find the
monopole density we need the probability distribution for the six
$F$ fields on the faces of a cube. The outwardly directed `plaquettes'
are labelled using the dice convention, namely that $F_n$ and $F_{7-n}$
are on opposite faces. We find
\begin{equation}
\begin{array}{ll}
\displaystyle
\Psi(F_1, \ldots ,F_6) &
\displaystyle         = \pi^{- \frac{5}{2}} 6^{\frac{1}{2}}
                      (a-b)^{\frac{3}{2}} (a+b) \beta^{\frac{5}{2}}
                      \delta(F_1 + \cdots +F_6) \\ [0.8em]
\displaystyle
                 & \displaystyle
                   \times {\rm exp}\{ -\beta a (F_1^2 + \cdots +F_6^2)
                      -2 \beta b (F_1 F_6 + F_2 F_5 + F_3 F_4) \} .
\end{array}
\label{psi6}
\end{equation}
This is the most general Gaussian form consistent with cubic symmetry
and the Bianchi identity. (Because $F_1 +\cdots +F_6=0$ we can
 add an arbitrary multiple of $(F_1+ \cdots +F_6)^2$ to the exponent
 without changing the distribution $\Psi$ at all. This freedom has
 been used to eliminate terms of the form $F_1 F_2$ etc.)
 The parameters $a$ and $b$ are fixed by the
known expectation values
\begin{equation}
\begin{array}{lll}
\displaystyle
\langle F_1^2 \rangle & \displaystyle
                       = \frac{2}{\beta} \int \frac{d^4 k}{(2\pi)^4}
  \frac{1-c_1}{4 - c_1 -c_2 -c_3 -c_4} & \displaystyle
                                         = \frac{1}{2\beta} ,\\ [0.8em]
\displaystyle
\langle F_1 F_2 \rangle & \displaystyle
                         = \frac{1}{2\beta} \int \frac{d^4 k}{(2\pi)^4}
\frac{-1+c_1+c_2-c_1c_2}{4-c_1-c_2-c_3-c_4} & \displaystyle
                                         = -\frac{\gamma}{2\beta} , \\
[0.8em]  \displaystyle
\langle F_1 F_6 \rangle & \displaystyle
                         = \frac{2}{\beta} \int \frac{d^4 k}{(2\pi)^4}
  \frac{-c_1+c_1c_2}{4-c_1-c_2-c_3-c_4} & \displaystyle
                                         = \frac{4\gamma -1}{2\beta} ,
\end{array}
\end{equation}
where $c_{\mu} = {\rm cos}k_{\mu}$. (On a finite lattice the integrals
are to be replaced by sums over allowed non-zero momenta.) On an
infinite lattice
\begin{equation}
\gamma = 0.215\,563 \ldots\, .
\end{equation}
  (Note that the value of $\gamma$ is calculated from the correlation
 between two $F$ fields, which is a gauge invariant quantity. Therefore
 $\Psi$ is gauge invariant.)
To give the correct expectation values, $a$ and $b$ must take the values
\begin{equation}
\begin{array}{lll}
\displaystyle
a & \displaystyle
   = \frac{1}{12} \frac{1+ \gamma}{\gamma (1- 2\gamma)} & \displaystyle
                       = 0.826\,049\ldots\, , \\
[0.8em]    \displaystyle
b & \displaystyle
   = \frac{1}{12} \frac{1- 5\gamma}{\gamma (1- 2\gamma)} & \displaystyle
                               = -0.052\,879\ldots\, .
\end{array}
\end{equation}
\begin{figure}[b]
\vspace{8cm}
\noindent \special{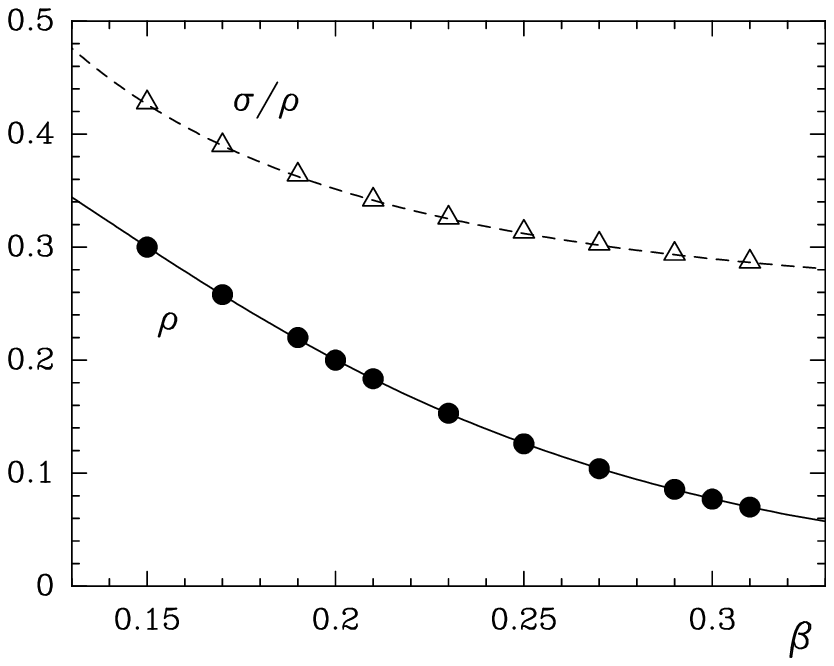}
 \hspace{2cm}\parbox{12cm}{\caption{The monopole density $\rho$
 and the $\sigma$
 to $\rho$ ratio as a function of $\beta$ for quenched QED. The symbols
 represent the data from ref.[20] together with our data, while
 the curves show the analytic results on an infinite lattice.}}
 \end{figure}
The monopole density $\rho(\beta)$ is
\begin{eqnarray}
\begin{array}{ll}
\displaystyle
\rho(\beta) & \displaystyle
             = \langle |M_{\mu}| \rangle \\ [0.8em]
\displaystyle
            & \displaystyle
             = \int_{-\infty}^{\infty} d F_1 \cdots d F_6 \,
             \Psi(F_1, ... ,F_6) \, |N_1 + \cdots + N_6| .
\end{array}
\end{eqnarray}

The fact that $|M_{\mu}|$ is bounded is enough to show that all
derivatives of the monopole density $\rho (\beta)$ are finite at all
$\beta$ values. If $\rho (\beta)$ is expanded as a series of the form
\begin{equation}
\rho (\beta) = \beta^{\frac{5}{2}} \sum_{n=0}^{\infty} a_n
(\beta_0 - \beta)^n
\label{reihe}
\end{equation}
about an arbitrary point $\beta_0$, then the bound on $|M_{\mu}|$ leads
to bounds on the $a_n$:
\begin{equation}
0 < a_n < \frac{1}{3} \frac{(2n + 3)!}{n! (n + 1)!} \frac{1}{4^n}
\beta_0^{-\frac{5}{2} - n} .
\end{equation}
These bounds are strong enough to show that the series in
eq.~(\ref{reihe})
is convergent with a radius of convergence of (at least) $\beta_0$. A
convergent series expansion rules out the existence of any essential
singularities in $\rho$. There is certainly no sign of a phase
transition in $\rho(\beta)$. A similar proof holds for correlation
functions involving a finite number of $f$'s and $N$'s.

\begin{figure}[b]
\vspace{13.5cm}
\noindent \special{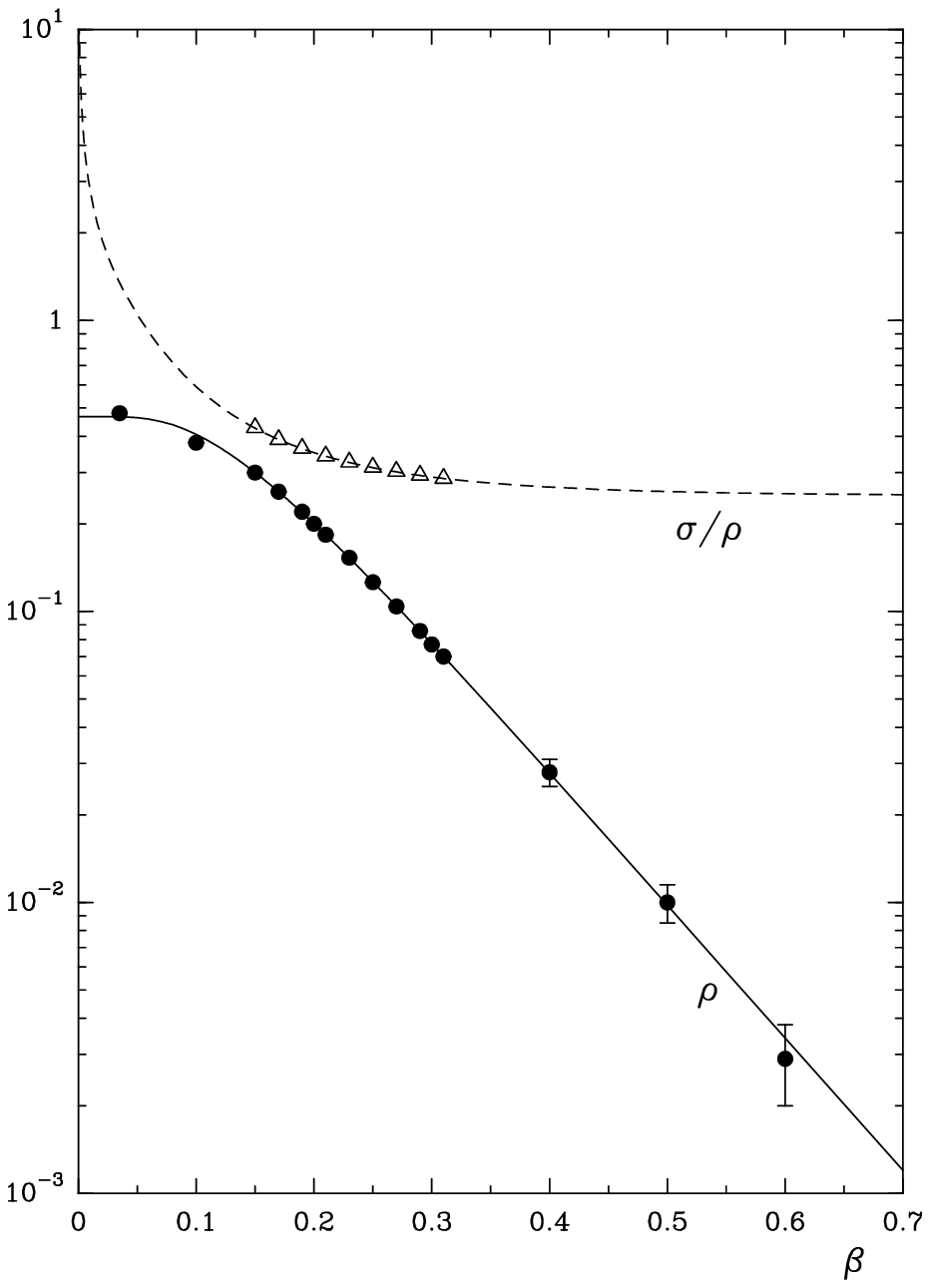}
 \hspace{2cm}\parbox{12cm}{\caption{   The same as
 fig.~2 for a larger range of $\beta$ on a
 logarithmic scale. The data points at the smaller and larger $\beta$
values are our measurements.}}
\end{figure}
In fig.~2 our formulae are checked against the Monte Carlo data~\cite{H&W}.
The agreement is excellent. In fig.~3 we show the monopole density
over a larger $\beta$ range. This enables us to show the asymptotic
limits of our formulae. At $\beta = 0$, $\rho(\beta)$ goes to 7/15,
while for large $\beta$ the density drops off like
$6 \,{\rm erfc} (\pi\beta^{\frac{1}{2}})$,
 i.e. approximately exponentially. Also in this
extended range we find good agreement with the data. The ratio
$\sigma(\beta)/\rho(\beta)$ goes to 1/4 at large $\beta$. This must be
so because at large $\beta$ the only topological excitations that
occur are isolated plaquettes with $|F| > \pi$, and these are
surrounded by a monopole loop of length 4. At small $\beta$ the ratio
diverges like $\beta^{-\frac{1}{2}}$ (cf. fig.~3). In the interesting
 region around
$\beta = 0.24$, $\sigma(\beta)/\rho(\beta)$ has only grown to
$\approx 0.34$, indicating that monopoles and antimonopoles are on
a short leash (i.e. are tightly bound). If we look at a timeslice
the average length of string
joining each monopole-antimonopole pair is simply $4\sigma/\rho$,
and is only about $1.4$ lattice units.

 We now turn to the discussion of the susceptibilities.
 To calculate these we need to know the general two-`plaquette'
 distribution $\Psi(F_i,F_j)$ and so the compact photon propagator
 $\langle f_i f_j \rangle $. If
 \begin{equation}
  \langle F_i^2\rangle = \langle F_j^2 \rangle = \frac {1}{2\beta}
  { \rm \ \ and\ \ \ }\langle F_i F_j \rangle = \frac{p_{ij}}{2\beta}
 \end{equation}
with $p_{ij}$ given by the non-compact photon propagator~\cite{propag},
\begin{eqnarray}
   \langle F_{\alpha \beta}(0) F_{\mu \nu}(x) \rangle &=&
                     \frac{1}{2\beta} \int \frac{d^4 k}{(2\pi)^4}\,
 \frac{1}{4-c_1-c_2-c_3-c_4}\  {\rm e}^{i k \cdot x} \times\nonumber \\
 & & \ \ \ \left(
   \delta_{\alpha \mu}
        (1-{\rm e}^{ -i k_\beta}) (1-{\rm e}^{ i k_\nu})
   - \delta_{\alpha \nu}
   (1-{\rm e}^{ -i k_\beta}) (1-{\rm e}^{ i k_\mu})\right. \nonumber \\
 & &\ \ \  \left.  - \delta_{\beta \mu}
        (1-{\rm e}^{ -i k_\alpha}) (1-{\rm e}^{ i k_\nu})
   + \delta_{\beta \nu}
     (1-{\rm e}^{ -i k_\alpha}) (1-{\rm e}^{ i k_\mu})  \right) \ \
   \end{eqnarray}
 then
 \begin{equation}
 \Psi(F_i,F_j)
   = \frac{\beta}{\pi} (1-p_{ij}^2)^{-\frac{1}{2}}
        \exp \{
    -\frac{\beta}{1-p_{ij}^2} (F_i^2+F_j^2)
                 +\frac{2 \beta p_{ij}}{1-p_{ij}^2} F_i F_j
                      \} .
 \label{psi2}
 \end{equation}

  With the help of the Fourier series
\begin{equation}
  f = -2 \sum_{k=1}^{\infty}\frac{(-1)^{k}}{k} \sin (k F)
\end{equation}
 we derive the series
\begin{equation}
 \langle f_i f_j \rangle = 4 \sum_{n=1}^{\infty} \sum_{m=1}^{\infty}
        \frac{(-1)^{n+m}}{n m} \,
       {\rm sinh}\left(\frac{n m p_{ij}}{2 \beta}\right)
    \exp \left( -\frac{n^2+m^2}{4 \beta} \right)
\label{dubsum}
\end{equation}
 for the compact propagator. Knowing $\langle f_i f_j \rangle$ for
 any pair of plaquettes we can calculate $\chi_m$ from its
 definition  eq.~(\ref{chif}).
\begin{figure}[t]
\vspace{13.5cm}
\noindent \special{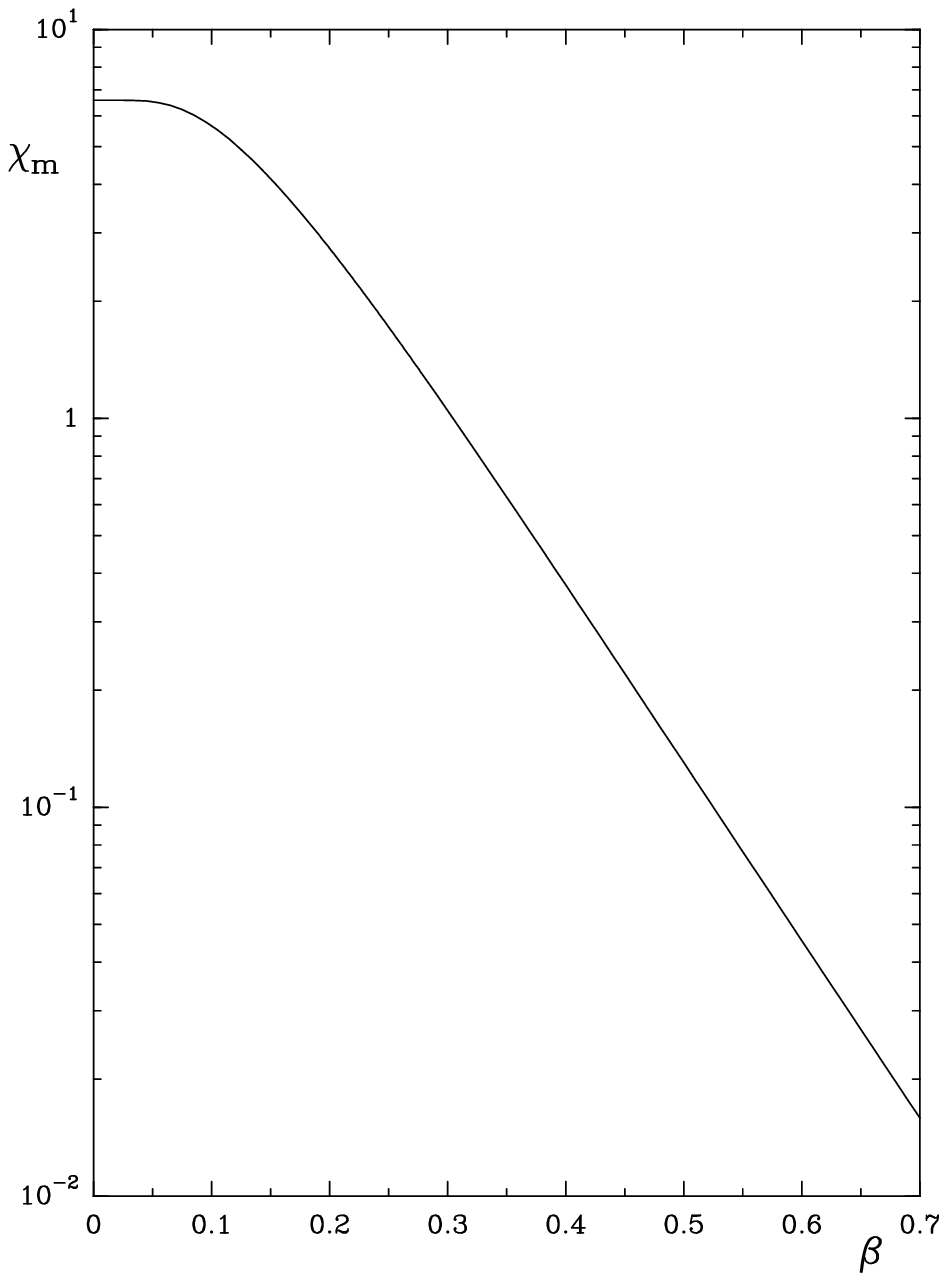}
\hspace{2cm}\parbox{12cm}{\caption {The monopole
  susceptibility $\chi_m$ for quenched QED as a
 function of $\beta$ on an infinite lattice.}}
\end{figure}

 In fig.~4 we
show the theoretical curve for $\chi_m$ as a function of $\beta$. At
finite $\beta$ there is no divergence: $\chi_m(\beta)$ is a correlation
function involving $f$'s (or $N$'s)
for which the proof of analyticity applies.
At $\beta=0$, $\chi_m$ has the value $\frac{2}{3}\pi^2$ because the
 $f$ fields are completely uncorrelated and evenly
 distributed in $(-\pi,\pi]$.
 It is striking that after $\beta \approx 0.1$ the curve
 drops exponentially in $\beta$ because the monopole density is
 dropping so quickly.

If we had monopole condensation we would expect that the monopole correlation
function ($x_4 \equiv t$)
\begin{equation}
 C(t) = \sum_{\vec{x}} \langle M_{\mu}(\vec{x},x_4)
 M_{\mu}(0)  \rangle
\end{equation}
 would exhibit long-range order~\cite{London}.
 $C(t)$ can be evaluated using the general form
  (\ref{dubsum}) for
 $ \langle f_i f_j \rangle$.
 In fig.~5 we show $C(t)$ on an infinite
lattice at the $\beta$ value ($\beta=0.244$) where Koci\'c et al.~\cite{Ko5}
place a phase transition. We find that the correlation function
drops by approximately three orders of magnitude between $t = 1$ and $2$.
\begin{figure}[b]
\vspace{13.5cm}
\noindent \special{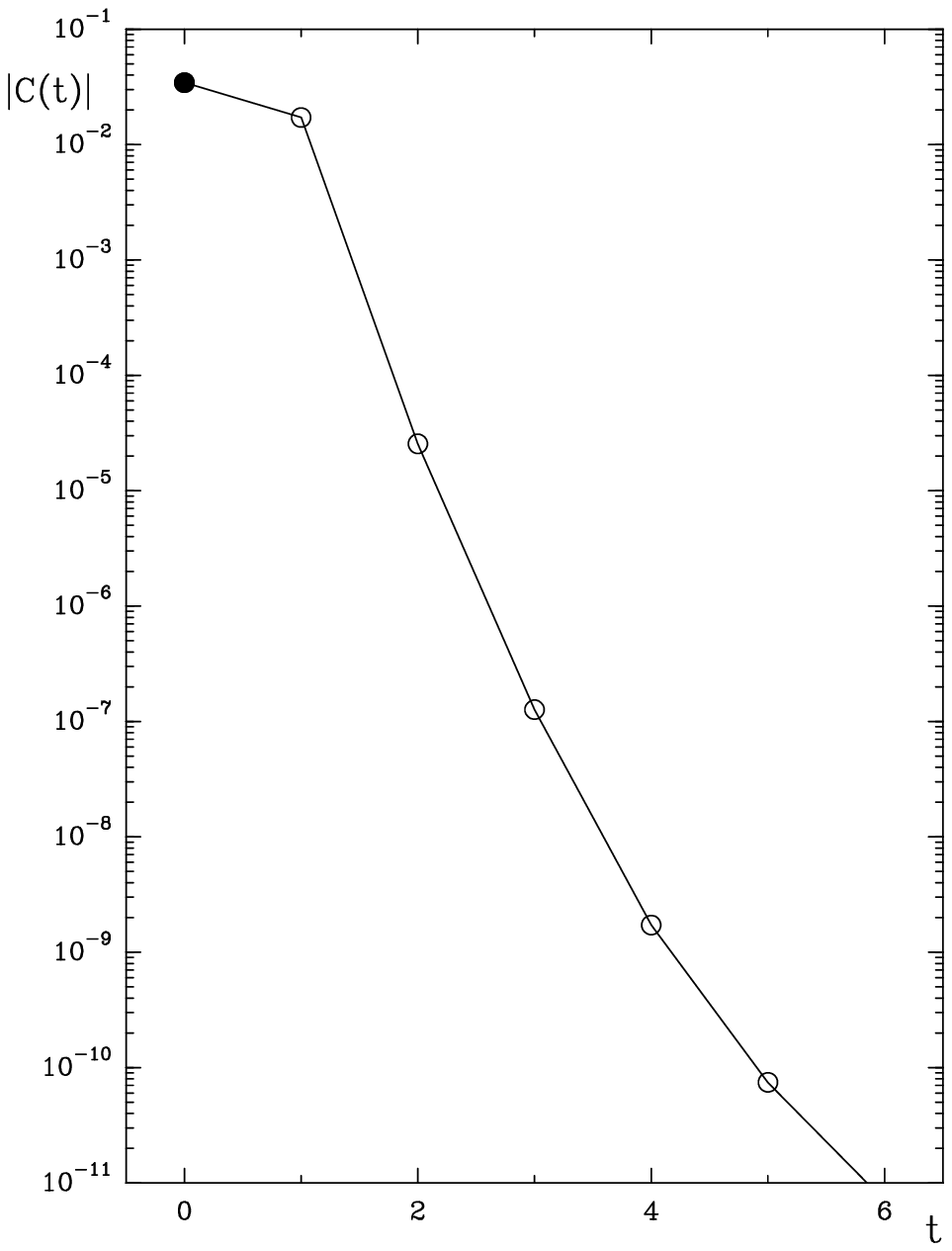}
 \hspace{2cm}\parbox{12cm}{\caption {The analytic result
 for the monopole correlation function
 $C(t)$ for quenched QED at $\beta = 0.244$ on an infinite lattice. The
 solid (open) symbols represent positive (negative) values of $C(t)$.}}
\end{figure}

\begin{figure}[t]
\vspace{8cm}
\noindent \special{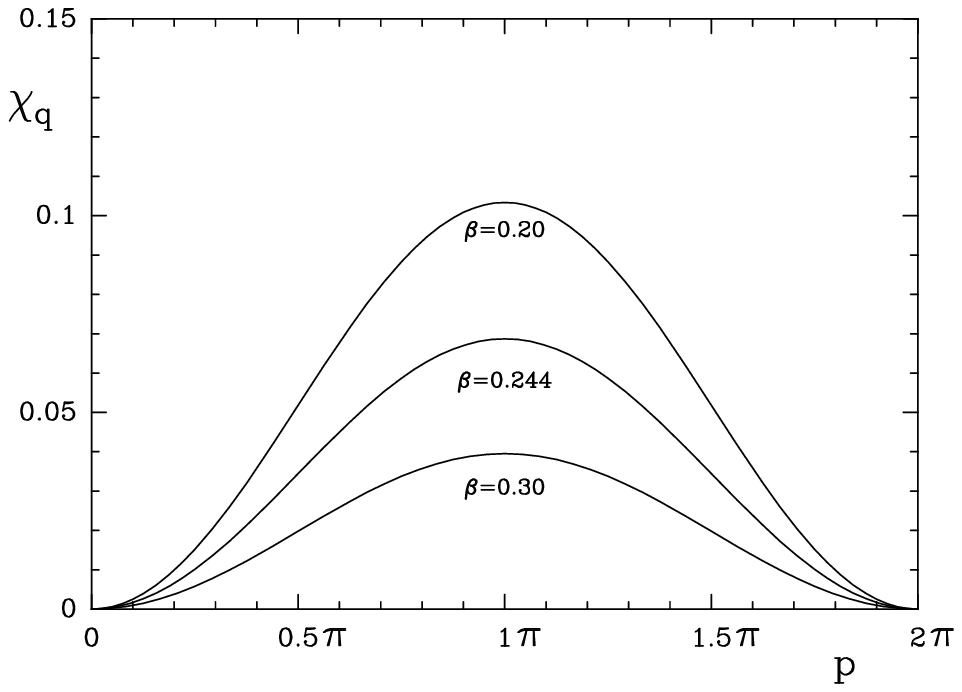}
\hspace{2cm}\parbox{12cm}{\caption {The susceptibility
 $\chi_q (p)$ as a function of momentum on an
infinite lattice for three values of $\beta$ for quenched QED. Koci\`c et al.
place a phase transition at $\beta = 0.244$. The momentum was chosen
along a lattice axis.}}
\end{figure}
The Fourier transform of $C(t)$ gives the charge susceptibility
$\chi_q(p)$ (see eq.~(\ref{chiq})) when the momentum $p$ is directed
 along a lattice axis.
 In fig.~6 we show $\chi_q(p)$ as a function of $p$ for
three different $\beta$ values. Because we can work on an infinite volume
we can go all the way to $p = 0$. We never see the enhanced long
 wavelength fluctuations characteristic of Bose-Einstein condensation
 (see fig.~1), on the contrary long wavelength fluctuations are
 strongly suppressed.
 The charge susceptibility does not diverge
 but vanishes like $p^2$ for all $\beta$ values. As usual the finite
 size effects are very small, and the curve measured on a finite lattice
 is already very close to that seen on an infinite lattice.

Many other quantities are also calculable. Particularly simple to
calculate are the compact Wilson loops, which again are analytic
and give a potential between charges which is Coulombic at all
$\beta > 0$, inconsistent with confinement.

The authors of ref.~\cite{Ko3} claim that strongly coupled dynamical
non-compact QED undergoes monopole condensation and confines electric
charge, the only evidence they present being the existence of a
percolation threshold. Their case is greatly weakened by the fact that
the quenched theory has a similar percolation threshold~\cite{Ko5}
(in fact, the evidence for the divergence of the Hands and Wensley
cluster susceptibility is strongest for the quenched theory), but
 as we have seen there is no monopole condensation or confinement.

 \newpage

\section{The dynamical case}
%\label{pot}

\begin{figure}[b]
\vspace{13.5cm}
\noindent \special{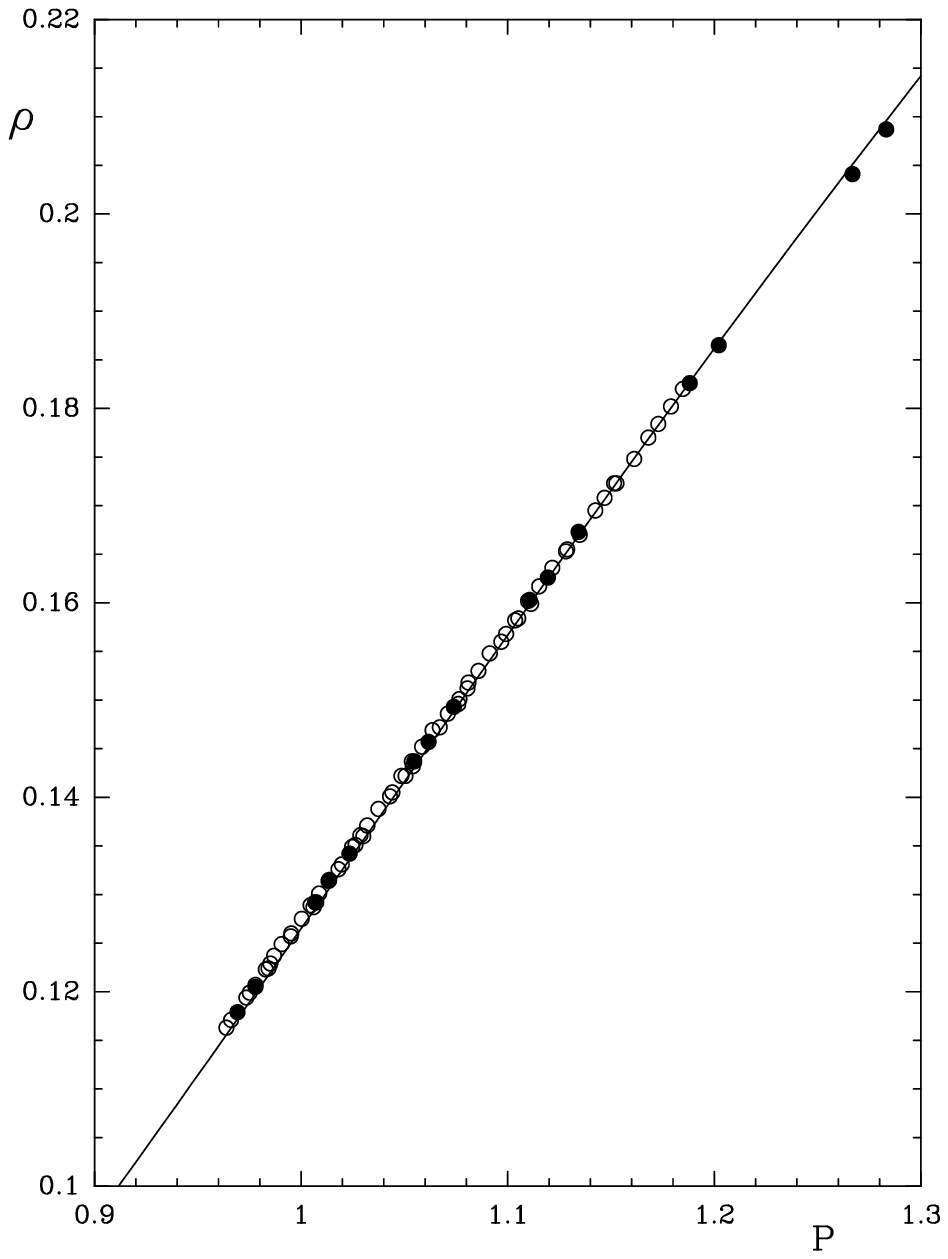}
 \hspace{2cm}\parbox{12cm}{\caption{The monopole
 density as a function of the plaquette energy.
 The data are for four flavours of dynamical staggered fermions. The solid
symbols represent our data, while the open symbols represent data from
ref.[19]. The curve is the analytic result for the quenched case
on an infinite lattice.}}
\end{figure}
We shall now investigate what happens if dynamical fermions are
included. For the action, which corresponds to four fermion
flavours, see eqs. (2.1-2.3).

If, as seems likely from the last section, the monopole properties are
determined by very short distance fluctuations of the electromagnetic
fields, we could expect that these properties are determined by the
plaquette energy values, because the plaquette energy is a good
measure of the fluctuation strength.

\begin{figure}[b]
\vspace{13.5cm}
\noindent \special{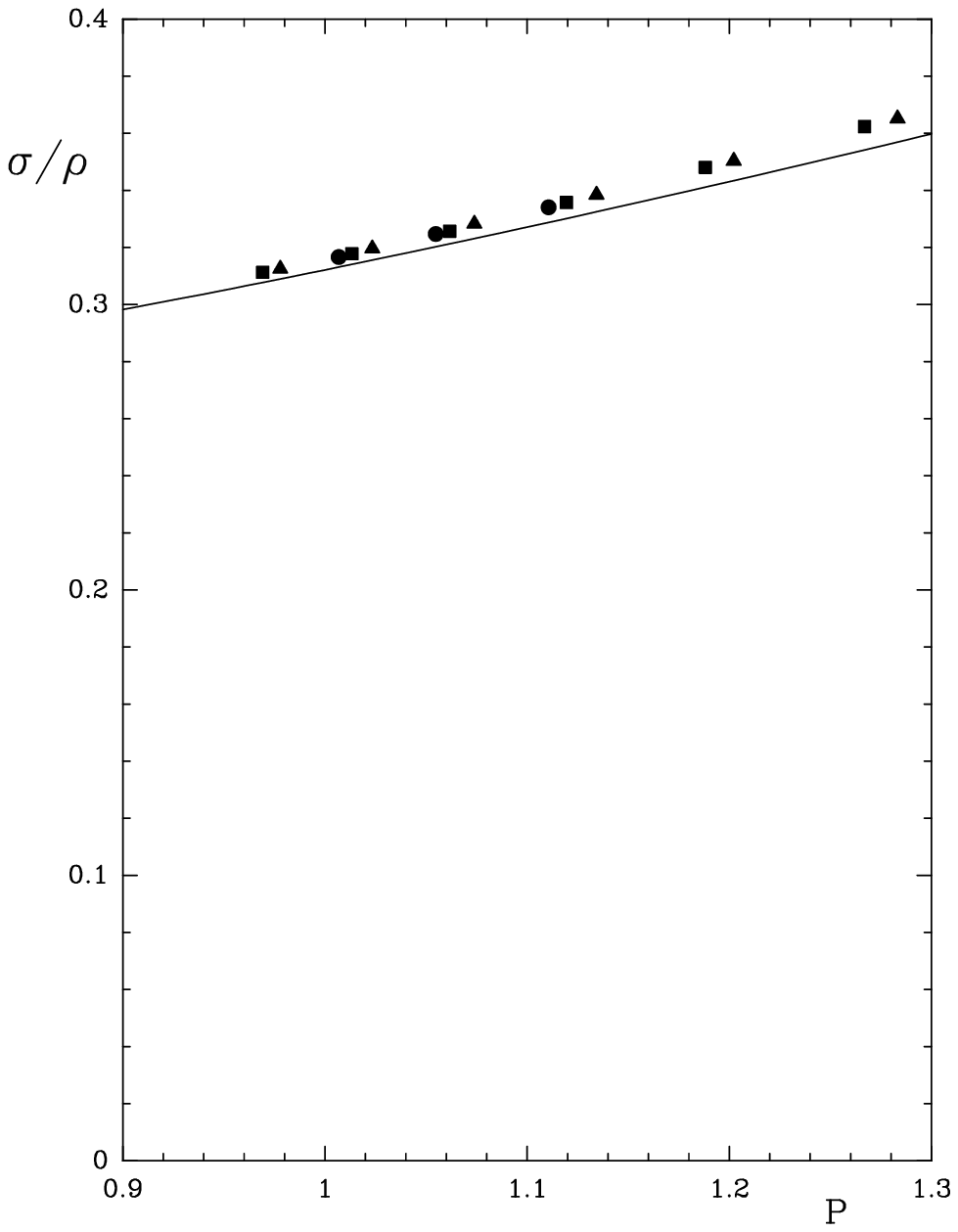}
 \hspace{2cm}\parbox{12cm}{\caption{The $\sigma$ to $\rho$ ratio
 as a function of the plaquette
 energy. The data are for four flavours of dynamical staggered fermions as
 given in Table~1. The triangle is $m = 0.04$, the square is
 $m = 0.02$ and the circle is $m =  0.01$. The curve is the analytic
 result for the quenched case on an infinite lattice.}}
 \end{figure}
Therefore we have plotted the monopole density against
$P \equiv \frac{1}{12} \sum_{\mu < \nu} \langle F_{\mu \nu}^2 \rangle$ in
fig.~7 using data from refs.~\cite{Ko7,GHRSS1} and
our results given in Table 1. We also show the analytic curve
calculated in the quenched case. We find surprisingly good agreement
between the data and the analytic result, indicating that the
inclusion of dynamical fermions does not change our previous
conclusions. The data comes from a wide range of bare masses,
$m = 0.005 - 0.16$, and plotting against $P$ has brought them all on
a universal curve. (Note that the quenched case can also be viewed
as the $m \rightarrow \infty$ limit.)

In fig.~8 we show the ratio $\sigma/\rho$. The measured values
are about 2\% higher than the quenched calculation. Plotting against
 $P$ has again brought measurements at different masses onto the same
 curve. The fact that $\sigma$ is a little different in the four
 flavour case than in the quenched case shows that the single
 $F$ distribution deviates slightly from a Gaussian form.

\begin{figure}[b]
\vspace{13.5cm}
\noindent \special{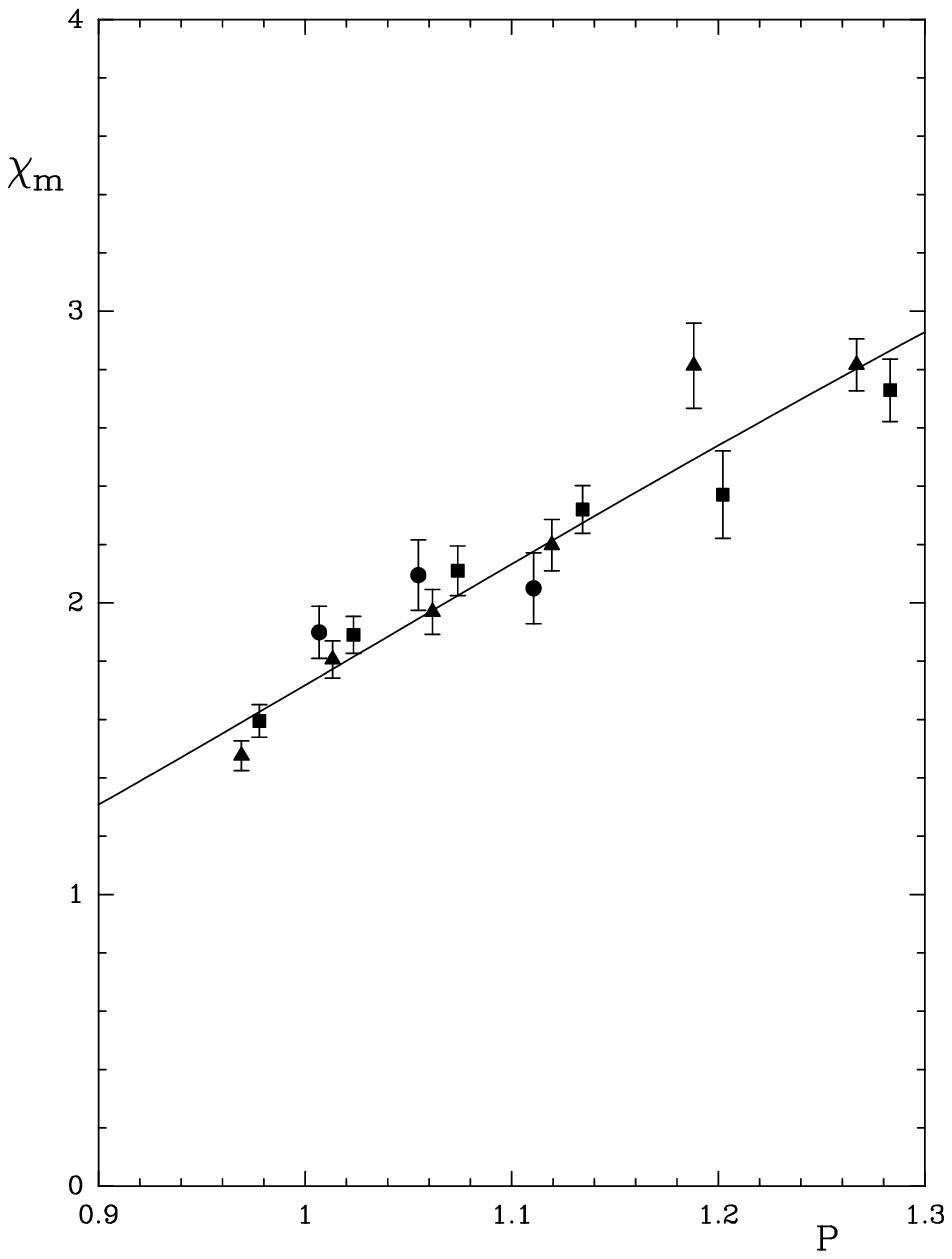}
 \hspace{2cm}\parbox{12cm}{\caption {The monopole susceptibility
 $\chi_m$ as a function of the
plaquette energy. The data symbols are the same as in the previous
figure. The curve is the analytic result for the quenched case on an
infinite lattice.}}
\end{figure}
The monopole susceptibility $\chi_m$ is plotted against $P$ in fig.~9.
Again, we compare the data with the analytic result.
 As has been remarked before~\cite{DeGrand},
this quantity is hard to measure accurately because
of large cancellations between positive and negative charges.
Within the errors we find agreement with the analytic quenched result,
and do not see any divergence of $\chi_m$. (The `phase transition'
reported in~\cite{Ko7} is at $P\approx 1.025$.)

\begin{figure}[b]
\vspace{8cm}
\noindent \special{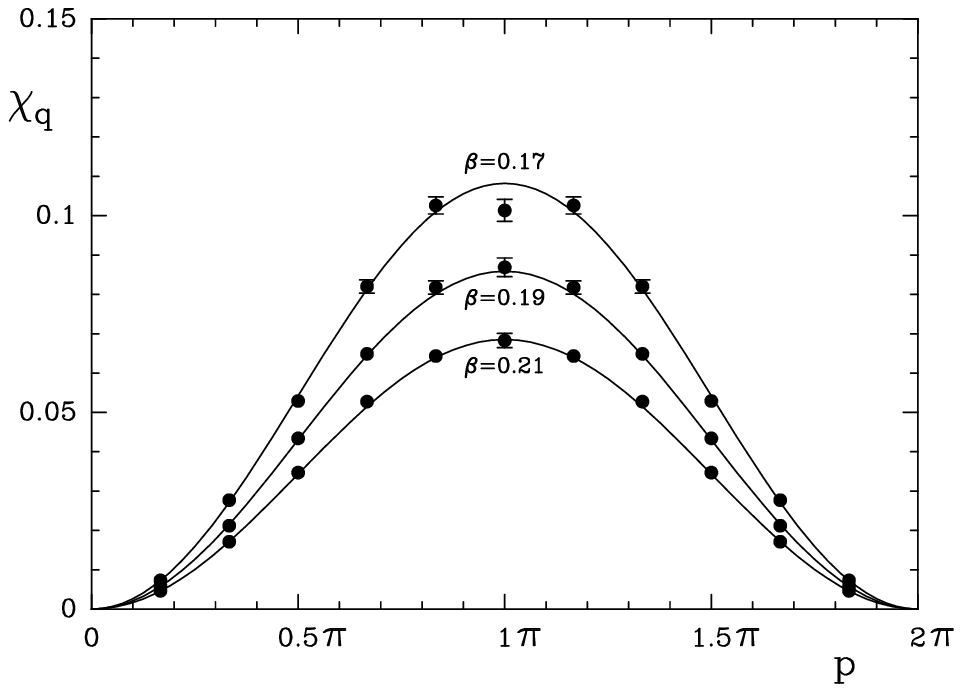}
\hspace{2cm}\parbox{12cm}{\caption {The susceptibility
 $\chi_q(p)$ as a function of momentum for
 three values of $\beta$.
 The data are for four flavours of dynamical staggered
fermions at $m = 0.04$. The curves are analytic results for the
quenched case on an infinite lattice.}}
 \end{figure}
  In fig.~10 we show $\chi_q(p)$ as measured on our four-flavour
configurations compared with curves calculated in the
quenched case at the same value of $P$. Measurement and calculation
are in excellent agreement. Even at small $\beta$ (the upper curve)
$\chi_q$ vanishes at small momenta, showing that the monopoles
 have not condensed. Consider for example $\chi_q$ measured at
the momentum $p=2\pi/L$. This measures charge density
 fluctuations with a
wavelength equal to the size of our lattice. Since a wave with
wavelength equal to the lattice size is positive in one half
of the lattice and negative in the other we are measuring the
 difference between the monopole concentrations in the two halves
 of our lattice. If the monopoles have truly condensed the tendency
 of bosons to occupy the same quantum state would cause
  large differences between the number found
in the right-hand and left-hand sides of our lattice.
 This happens for
a condensed boson gas (see fig.~1), but for monopoles we see that long
wavelength fluctuations are strongly suppressed, showing that
condensation has not taken place.

The surprising agreement between all the theoretical curves (with no
adjustable parameters) and the Monte
Carlo data confirms our hypothesis that the monopole properties are
determined by short-distance fluctuations.
%\vspace{0.5cm}

\begin{table}[h]
   \begin{center}
      \begin{tabular}{||l|l|l|l|l|l||}
         \hline
         \multicolumn{1}{||c|}{$\beta$} &
         \multicolumn{1}{c|}{$m$} &
         \multicolumn{1}{c|}{$P$} &
         \multicolumn{1}{c|}{$\rho$} &
         \multicolumn{1}{c|}{$\sigma$} &
         \multicolumn{1}{c||}{$\chi_m$} \\
         \hline
 $0.17$ & $0.04$ & $1.2832(6)$ & $0.2087(2)$ & $0.07621(9) $ & $2.729(107)$\\
 $    $ & $0.02$ & $1.2669(6)$ & $0.2041(2)$ & $0.07395(9) $ & $2.816(89) $\\
 \hline
 $0.18$ & $0.04$ & $1.2022(6)$ & $0.1865(3)$ & $0.06533(13)$ & $2.371(150)$\\
 $    $ & $0.02$ & $1.1881(6)$ & $0.1826(3)$ & $0.06355(12)$ & $2.813(146)$\\
 \hline
 $0.19$ & $0.04$ & $1.1343(5)$ & $0.1673(2)$ & $0.05662(7) $ & $2.320(82) $\\
 $    $ & $0.02$ & $1.1194(5)$ & $0.1626(2)$ & $0.05459(7) $ & $2.198(88) $\\
 $    $ & $0.01$ & $1.1106(9)$ & $0.1603(3)$ & $0.05355(11)$ & $2.050(122)$\\
 \hline
 $0.20$ & $0.04$ & $1.0739(3)$ & $0.1493(2)$ & $0.04901(7) $ & $2.110(85) $\\
 $    $ & $0.02$ & $1.0617(4)$ & $0.1457(2)$ & $0.04745(6) $ & $1.969(77) $\\
 $    $ & $0.01$ & $1.0548(6)$ & $0.1437(2)$ & $0.04666(10)$ & $2.095(121)$\\
 \hline
 $0.21$ & $0.04$ & $1.0234(4)$ & $0.1342(2)$ & $0.04289(6) $ & $1.890(63) $\\
 $    $ & $0.02$ & $1.0133(4)$ & $0.1314(2)$ & $0.04176(6) $ & $1.806(64) $\\
 $    $ & $0.01$ & $1.0068(4)$ & $0.1292(2)$ & $0.04091(9) $ & $1.899(89) $\\
 \hline
 $0.22$ & $0.04$ & $0.9779(3)$ & $0.1205(2)$ & $0.03766(6) $ & $1.595(56) $\\
 $    $ & $0.02$ & $0.9692(3)$ & $0.1179(2)$ & $0.03670(5) $ & $1.476(51) $\\
         \hline
      \end{tabular}
   \end{center}
   \hspace{2cm}\parbox{12cm}{\caption[xxx]{The monopole
 density $\rho$, the string density
$\sigma$ and the monopole susceptibility $\chi_m$ on a $12^4$ lattice
with four flavours of dynamical staggered fermions. Also given are the
plaquette energy values $P=\frac{1}{12} \sum_{\mu < \nu} \langle
F^2_{\mu \nu} \rangle$.}}
    \label{table1}
\end{table}

%\newpage

\section{Discussion}

In this work we have investigated monopoles in quenched and dynamical
non-compact lattice QED. In the quenched case we have derived analytic
formulae which we have checked against numerical data. Here we can
prove that there are no singularities and divergences in the quantities
we have looked at. The same formulae describe the dynamical case
when quantities are plotted against the plaquette energy, which measures
the strength of the electromagnetic field. Thus we have arrived at a
quantitative understanding of monopoles in both the quenched and
dynamical case.

The similarity of the percolation threshold in the
quenched, $N_f = 2$ and $N_f = 4$ cases~\cite{Rakow4}, occurring
at almost the same monopole density and with almost identical critical
exponents, lends further support to the idea that the monopole
behaviour is the same in the quenched and dynamical case. This picture
may, however, change when $N_f$ is so large that the phase transition
becomes first order~\cite{largeN}.

We have also looked at the distribution of monopoles and antimonopoles
in a time slice. We saw that 60$\%$ of all monopoles have an
antimonopole on adjacent sites and only 10$\%$ a monopole. So actually
what we see looks more like a gas of dipoles than a condensate of
monopoles. (One can also see this from the low $\sigma/\rho$ value;
see figs. 2,3, 8.)

In dynamical QED we have already checked~\cite{GHRSS1} that the
potential is Coulombic and the photon does not acquire a mass. This is
also inconsistent with confinement at low $\beta$.

It is not surprising that when the monopole density becomes large
($\rho \gap 0.15$) percolation takes place. However, percolation is
not necessarily connected
with condensation or with any other field-theoretic
or thermodynamic property of the theory. Indeed, it is rather easy to
find examples where the percolation threshold and the ``authentic''
phase transition are at different couplings. One example is the
Ising model of higher dimension, where the percolation threshold lies
at higher $\beta$ than the phase transition~\cite{Aizenman}. Also, if
one looks at the 3d Ising model at $\beta = 0$ and non-zero
magnetic field, $h$, one finds percolation thresholds at certain values
of $h$, because if $\beta = 0$ the spins are randomly distributed
with the concentration controlled by the magnetic field. It is well
known that randomly distributed sites on a cubic lattice first
percolate when the concentration has reached $\approx 32\%$
\cite{Essam}. So for strong negative $h$ the up-spins do not percolate,
while the down-spins do. There is a percolation threshold at about
32$\%$ ($h\approx -0.38$) and then both spins percolate. At a concentration
of $\approx 68\%$ ($h\approx + 0.38$) there is a second threshold above
 which only the up-spins percolate. Despite the occurrence of these
 percolation thresholds there are
 certainly no phase transitions at non-zero $h$ in the Ising model.

The cluster susceptibility of Hands and Wensley~\cite{H&W} is
\begin{equation}
\chi_c = \langle \frac{(\sum_{n=4}^{n_{max}} g_n n^2) - n_{max}^2}
{\sum_{n=4}^{n_{max}} g_n n} \rangle,
\label{chiclust}
\end{equation}
where $n$ is the number of dual sites in a cluster linked together by
monopole world lines, $g_n$ is the number of clusters of size $n$
and $n_{max}$ is the size of the largest cluster. This susceptibility
has long been used to find percolation thresholds (it is essentially
the $S(p)$ of~\cite{Essam} or the $\chi^f$ of~\cite{Grimmett}).
However it can lead to
 misleading results if it is used to locate phase transitions
due to the following defects. First of all,
$\chi_c$ is not a Green's function, so a divergence of $\chi_c$ does
not imply an infinite correlation length and so does not indicate a
second order phase transition. Furthermore, $\chi_c$ counts monopoles
and antimonopoles with the same sign, whereas physically they should
contribute with opposite sign.

Moreover, in order to compute $\chi_c$ one must treat monopoles
differently depending on whether or not they belong to the same cluster.
No physical operator can do this because by Bose symmetry all physical
operations treat indistinguishable particles identically. If
$\chi_c$ is not a physically realizable quantity the fermions cannot
couple to it and the divergence of $\chi_c$ cannot cause a chiral
phase transition.

It is worth noticing that in dynamical non-compact QED $\chi_c$
diverges in places where there is no phase transition. The chiral phase
transition takes place only at $m = 0$, while at finite $m$
quantities such as $\chibarchi$ are smooth functions of $\beta$.
 However, $\chi_c$ diverges not only at
$m = 0$ but for all $m$ including $m = \infty$, (see
ref.~\cite{Ko6,Ko7}).

In conclusion, the papers [12-20] on monopoles in non-compact QED
do not prove that monopoles are relevant in the continuum limit
of the lattice theory, and so do not invalidate the picture of the
chiral phase transition presented in refs. [1-11].
\vspace{0.5cm}

\section*{Acknowledgements}
\label{acknowledgements}

This work was supported in part by the Deutsche Forschungsgemeinschaft.
We furthermore like to thank E. Seiler for drawing our attention to
ref.~\cite{Aizenman}.

\newpage

\bibliographystyle{unsrt}

\clearpage

\end{document}